\documentclass[12pt]{iopart}
\usepackage{CJK}
\usepackage{graphicx}
\usepackage{epsfig}
\usepackage{url}
\usepackage{amssymb}
\def\beq{\begin{equation}}
\def\eeq{\end{equation}}
\def\beqar{\begin{eqnarray}}
\def\eeqar{\end{eqnarray}}

\def\isotope#1#2{\mbox{${}^{#2}{\rm #1}$}}
\def\fe5#1{\isotope{Fe}{5#1}}
\def\co5#1{\isotope{Co}{5#1}}
\def\ni5#1{\isotope{Ni}{5#1}}

\def\apj{ApJ}
\def\aap{A$\&$A}
\def\apjl{ApJL}

\def\nat{Nature}

\def\aj{The Astronomical Journal}

\def\mnras{MNRAS}
\def\prd{PhRvD}

\def\prc{PhRvC}
\def\nphysa{NuPhA}
\def\araa{ARA$\&$A}

\def\fun#1#2{\lower3.6pt\vbox{\baselineskip0pt\lineskip.9pt
 \ialign{$\mathsurround=0pt#1\hfil##\hfil$\crcr#2\crcr\sim\crcr}}}

\begin{document}
%\begin{CJK*}{GBK}{ }

\title[Spallation of \textit{r}-Process Nuclei]{Spallation of \textit{r}-Process Nuclei Ejected from a Neutron Star Merger}

%\author{Xilu Wang (王夕露)$^{1,2}$, Brian D. Fields$^{3,4}$, Matthew Mumpower$^{5,6}$, Trevor Sprouse$^{2,5}$, Rebecca Surman$^{2}$, Nicole Vassh$^{2}$}
\author{Xilu Wang$^{1,2}$, Brian D. Fields$^{3,4}$, Matthew Mumpower$^{5,6}$, Trevor Sprouse$^{2,5}$, Rebecca Surman$^{2}$, Nicole Vassh$^{2}$}

\address{$^{1}$Department of Physics, University of California, Berkeley, CA 94720, USA}
\address{$^{2}$Department of Physics, University of Notre Dame, Notre Dame, IN 46556, USA}
\address{$^{3}$Department of Astronomy, University of Illinois at Urbana-Champaign, Urbana, IL 61801, USA}
\address{$^{4}$Department of Physics, University of Illinois at Urbana-Champaign, Urbana, IL 61801, USA}
\address{$^{5}$Theoretical Division, Los Alamos National Laboratory, Los Alamos, NM, 87545, USA}
\address{$^{6}$Center for Theoretical Astrophysics, Los Alamos National Laboratory, Los Alamos, NM, 87545, USA}

\ead{xwang50@nd.edu, xlwang811@gmail.com}
%\nocollaboration

%\end{CJK*}

\vspace{10pt}
\begin{indented}
\item[]November 2019
\end{indented}

\begin{abstract}
Neutron star mergers (NSMs) are rapid neutron capture (\textit{r}-process) nucleosynthesis sites, which eject materials at high velocities, from $0.1c$ to as high as $0.6c$. Thus the r-process nuclei ejected from a NSM event are sufficiently energetic to initiate spallation reactions with the interstellar medium (ISM) particles. With a thick-target model for the propagation of high-speed heavy nuclei in the ISM, we find that spallation reactions may shift the r-process abundance patterns towards solar data, particularly around the low-mass edges of the \textit{r}-process peaks where neighboring nuclei have very different abundances. The spallation effects depend both on the astrophysical conditions of the r-process nuclei and nuclear physics inputs for the nucleosynthesis calculations and the propagation process. This work extends that of \cite{Wang2019} by focusing on the influence of nuclear physics variations on spallation effects.
\end{abstract}

\noindent{\it Keywords}: \textit{r}-process, nucleosynthesis, nuclear reaction cross sections, nuclear abundances, compact binary stars
%\maketitle

\maketitle

\section{Introduction}

The rapid neutron capture process (\textit{r} process) is one dominant nucleosynthesis avenue for heavy elements, especially for those heavier than the iron group \cite{Burbidge1957, Cameron1957}. In the \textit{r} process, rapid neutron capture pushes material far from stability and shapes the characteristic abundance pattern with three distinct peaks (at mass numbers $A\sim80$,  $A\sim130$, and $A\sim196$). These peaks are clearly seen in the abundance pattern of our solar system \cite{Lodders2003, Sneden2008}, where approximately half of the heavy elements have an \textit{r}-process origin. 

NSMs are one confirmed site of \textit{r}-process nucleosynthesis \cite{Abbott, NSM}, with the kilonova signal from the multi-messenger event GW170817 which indicated lanthanide production from a NSM \cite{ Kasen2017, Cowperthwaite2017}. The r-process nuclei ejected from NSMs are expected to travel with high speed that ranges from $0.1c$ to as high as $0.6c$, based on kilonova models \cite{Li1998, Tanaka2013, Kasen2017, Rosswog2018, Wollaeger2018}, and NSM simulations for a dynamical ejecta \cite{Bauswein2013, Hotokezaka2013, Rosswog2013, Endrizzi2016, Lehner2016, Sekiguchi2016, Rosswog2017} or a viscous and/or neutrino-driven wind \cite{Surman2008, Chen2007, Dessart2009, Wanajo2014, Just2015, Perego2014, Martin2015, Siegel2018}. 

What happens if these energetic heavy particles ejected from a NSM are traveling through the ISM? Obviously these particles would interact with the ISM and they are sufficiently energetic to initiate spallation: nuclear fragmentation processes in which a heavy nucleus emits one or more nucleons, thus reducing its atomic weight. This interaction is well studied in the context of cosmic rays. 
%Cosmic rays are about 5 orders of magnitude more enriched in Li, Be, and B than the ISM or solar system, due to fragmentation of cosmic ray C, N, and O nuclei during propagation in the interstellar medium \cite[e.g.,][]{George2009}. 
The effect of spallation on the cosmic-ray abundance pattern is to ``fill in the valley'' at Li, Be, and B, at the expense of a small reduction in the neighboring CNO peak \cite{CR1, Duncan1992, Fields1994, Higdon1998,  Lemoine1998, Reeves1970, Walker1985, Ramaty2000, Fields2000, Suzuki2001}. Thus, spallation reactions may also influence the overall \textit{r}-process nucleosynthesis yields from a NSM in a similar way. 

There are many uncertainties in \textit{r}-process nucleosynthesis, including nuclear inputs for unstable nuclei, and astrophysical conditions of the merger event \cite{Matt2016, Kajino2017}.  These uncertainties bring in the large variations around the second peak ($A\sim130$) and third peak ($A\sim196$) of \textit{r}-process abundance patterns, which generally don't match the solar data well. 
So in this paper, we investigate the effect of spallation on the shapes of the \textit{r}-process abundance peaks produced in fast ejecta from a NSM event, and test whether spallation could alleviate the mismatch between simulation results and solar data. In doing so, we explore the impact with different nuclear physics inputs and astrophysical conditions.

The next section will
briefly introduce the methods of our spallation calculation. Spallation results for \textit{r}-process abundances calculated with different astrophysical and nuclear physics inputs are presented in Section~\ref{sec: results}. In Section~\ref{sec: discussion}, further discussions
and conclusions are given. For detailed model construction and results, see \cite{Wang2019}. This paper mainly investigate the influence of nuclear physics variations on spallation effects, with comparison of different theoretical calculations for spallation cross-sections, compared with \cite{Wang2019}.

\section{Method}
\label{sec: method}

To investigate the potential influence of spallation on \textit{r}-process abundance patterns, we first generate initial abundance patterns of \textit{r}-process nuclei ejected from a NSM using the nucleosynthesis network code PRISM \cite{Matt2016, Matt2017, Matt}. We then adopt a thick-target model for propagation of the \textit{r}-process nuclei through the ISM, which assumes all the \textit{r}-process nuclei will finally interact with the ISM and the ionization loss dominates the total energy loss mechanism, obtaining new abundances of the \textit{r}-process nuclei after spallation. Detailed assumptions and calculations are found in \cite{Wang2019}.

\subsection{Equations}

We adopt the transport equation with a thick-target approximation \cite{Wang2018} for the propagation of the r-process nuclei :
\beq
\label{eq:prop}
\partial_t N_E \approx  \partial_E (b_E N_E)  \ + q_E  \ .
\eeq
Here and throughout, $E$ denotes {\em kinetic energy per nucleon in MeV}, which depends only on the relative velocity between the projectile and target and thus is the same viewed from either frame. The instantaneous number of propagated particles per energy per nucleon at time $t$ is $N_E(E,t)=dN/dE$, thus $N_E \ dE$ is the number of the propagated ejecta nuclei with kinetic energy in the range $(E,E+dE)$. The source function is $q_E=dN/dEdt$ and $b_E=-dE/dt$ is the rate of energy loss (per nucleon).
Velocity of the ejecta relative to the ISM is $v(E)=[1-(1+E/(m_pc^2))^{-2}]^{1/2}c$, $m_p$ is the proton mass, such that for $v(E)=0.3c$, $E\sim45.29$ MeV. We assume that the only important loss mechanism is the energy loss due to ionization and spallation reactions. 

We calculate a set of spallation reactions $i+j \rightarrow \ell + \cdots$ in which projectile $i$ and target $j$ nuclei give rise to products $\ell$, and the number fraction of the total spallation-produced nuclei ${\ell}$ at time $t_f$ to the initial projectile $i$ at time $t_0$ is:
\beqar
\label{eq: fraction}
f_{i}^{\ell}&=& \sum_{j} f_{i,j}^{\ell}=\sum_{j} \frac{N_{ij}^{\ell}(t_f)}{N_{i,0}}=\sum_{j} y_{j} \int_{E_x(t_f)}^{E_0} \frac{\sigma_{ij}^{\ell}(E') \, v(E') \, dE'} {b_{i,E'}(n_{\rm gas})/n_{\rm gas}} \ , 
\eeqar
where $E_{0}$ is the initial kinetic energy per nucleon of the projectile nuclei $i$, which is the maximum kinetic energy of the nuclei. $E_x(t_f)$ is the kinetic energy per nucleon of the projectile nuclei $i$ at time $t_f$ when the nuclei are no longer energetic enough to have spallation reactions.  Here $\sigma_{ij}^{\ell}(E)$ is the cross section for the production of nuclei ${\ell}$ by the spallation reaction between ejecta nuclei $i$ and ISM nuclei $j$. The weighting $y_j=n_j/n_{\rm gas}$ is the fraction by number of ISM particles in the form of $j \in ({\rm H,He})$, with $n_{\rm gas}$ is the total number density of ISM particles.

For each nucleus $i$ interacting with ISM nucleus $j$ through spallation reaction $i+j\to \ell + \cdots$, we calculate $f_{i,j}^{\ell}$ ($A_{\ell}=[A_i-n,A_i-0]$, $n \in (10,25)$). Because the particle number during propagation is conserved, 
nucleus $i$ (abundance $Y_i$) produces the same number of nucleus ${\ell}$ (abundance $Y_\ell$), thus the loss of nucleus $i$ during the propagation is at the same number as the production of all the nucleus from the spallation reaction of nucleus $i$, i.e., $f_{i,\rm prop\ loss}=\sum\limits_{\ell} (y_{p}f_{i,p}^{\ell}+ y_{\alpha}f_{i,\alpha}^{\ell})$. The new abundance after spallation is therefore
\beqar
\label{eq:change}
Y_{i,\rm spallation} 
& = & Y_i(1-f_{i,\rm prop\ loss})+\sum\limits_{k} (y_{p}f_{k,p}^iY_k)+\sum\limits_{k} (y_{\alpha}f_{k,\alpha}^iY_k) \ \ .
\eeqar
To compare the new abundance pattern with the initial \textit{r}-process abundance pattern, we compute the spallation abundance change ratio by
\beqar
\label{eq:change_ratio}
F_{i,\rm {change}}=(Y_{i,\rm spallation}-Y_i)/Y_{i}. 
\eeqar

\subsection{Spallation cross-sections}
\label{subsec: cross}

We need to know the cross sections for nuclear spallation in order to investigate the spallation effects, based on \Eref{eq: fraction}. 
However, there are little experimental data available for spallation reactions between a proton or $^{4}$He and a target nuclide which is heavier than iron, in the energy range smaller than $\lesssim100$ MeV. Therefore we adopt the theoretical spallation/inelastic cross sections from TALYS/1.9 \cite{TALYS,TALYS2}\footnote{https://tendl.web.psi.ch/tendl\_2019/tendl2019.html} with default nuclear inputs.

\begin{figure}[]
%\centering
\includegraphics[scale=0.48]{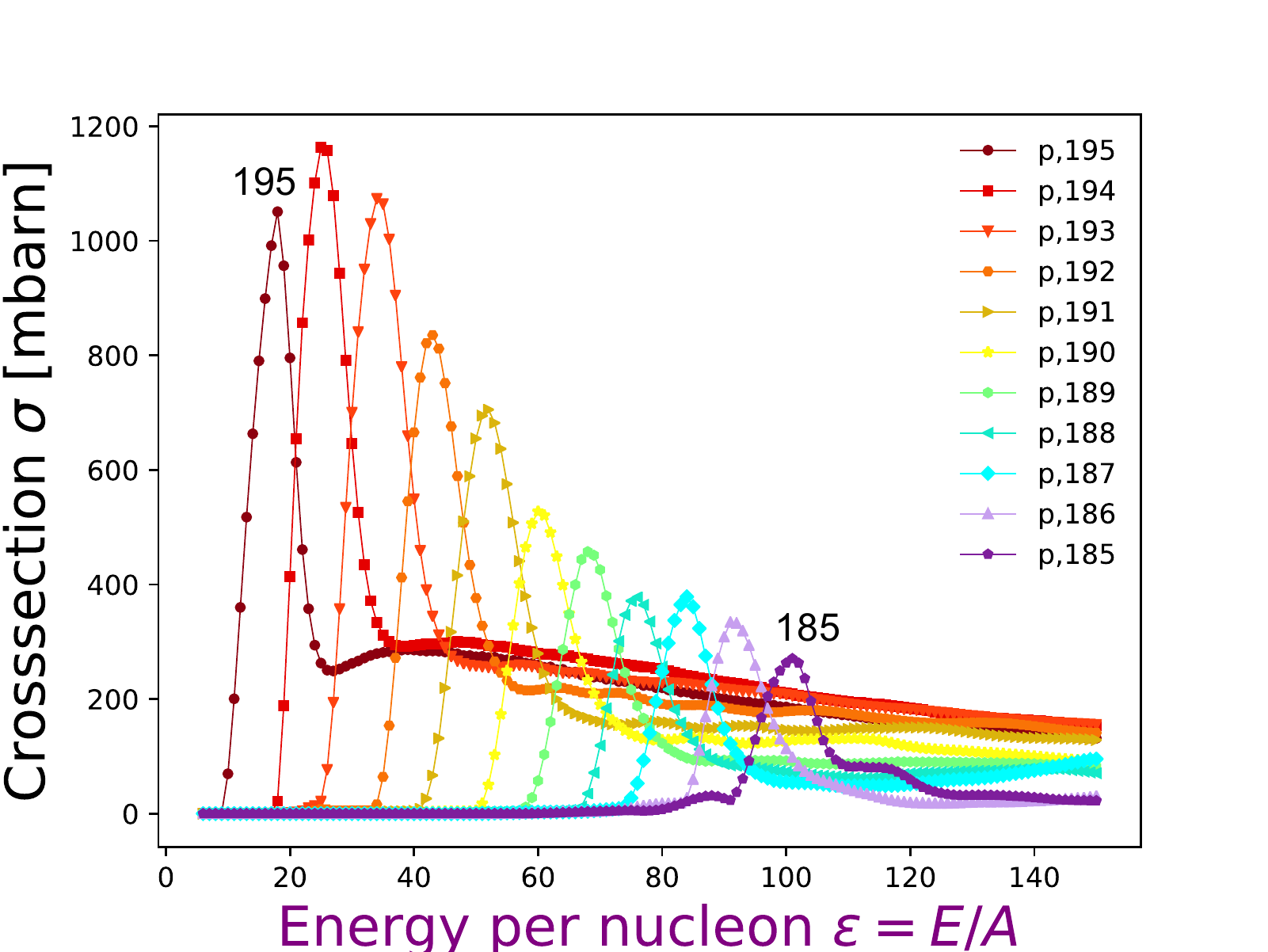}
\includegraphics[scale=0.48]{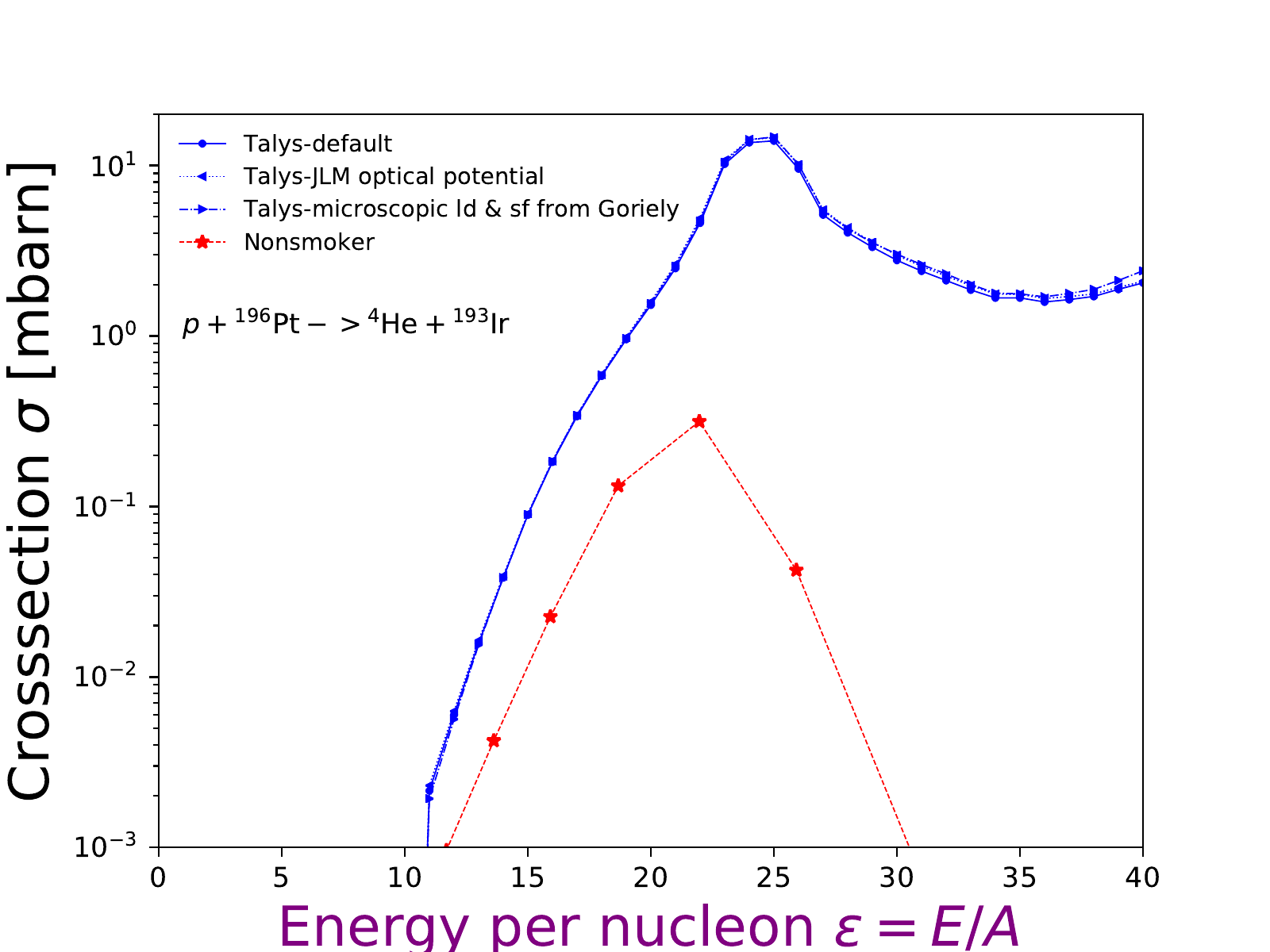}
\caption{
Left: Individual cross sections for each spallation channel for the reaction between $^{196}{\rm Pt}$ and a proton, generated with TALYS. The colored lines show the cross sections for the individual channels $^{196}\mathrm{Pt}(p,x)A$, where $A$ is the mass number of the final nucleus after spallation. As the projectile energy per nucleon increases, the dominant spallation production channel moves from $A\sim195$ (dark red circles) to smaller mass numbers, with a wide range of product nuclei at the highest energies.
Right: Comparison of two theoretical calculations from TALYS (default inputs: blue solid line; JLM microscopic optical model potential: blue dotted line; microscopic level densities from Goriely's table $\&$ Goriely's hybrid model for gamma-strength functions: blue dash-dot line) and NONSMOKER (red line) for the cross section values for an example individual spallation channel $^{196}\mathrm{Pt}(p,\alpha)^{193}\mathrm{Ir}$.
\label{fig:crosssection_channel}}
\end{figure}

Left panel in Figure~\ref{fig:crosssection_channel} shows the cross sections for each spallation channel $^{196}\mathrm{Pt}(p,x)$. The relevant energy range here is roughly $5-100$ MeV; moving from high projectile energy per nucleon to low within this range, the dominant creation channel shifts from producing $A=185$ to $A=195$ nuclei. Spallation reactions change a projectile nucleus to a new nucleus with nearby but smaller mass number.  Thus we would expect spallation to shift the \textit{r}-process abundance pattern peaks to smaller mass numbers, which is confirmed by the results presented in Section~\ref{sec: results}. 

Different theoretical calculations can give large variations in the spallation cross-section values. 
Right panel in Figure~\ref{fig:crosssection_channel} shows the comparison of the cross-section values generated for the spallation channel $^{196}\mathrm{Pt}(p,\alpha)^{193}\mathrm{Ir}$ by TALYS and NONSMOKER calculations \cite{NS1,NS2,NS3}\footnote{https://nucastro.org/nonsmoker.html}. TALYS contains a variety of options for input nuclear physics such as nuclear level densities, gamma-strength functions, and optical potentials. We compare the calculation results from default models with results from variations in available inputs, finding that the calculated cross-section values differ by at most 10 percent. However, the cross section value for the spallation channel obtained from NONSMOKER calculations differs from TALYS by more than an order of magnitude. 
Thus, we calculate spallation effects using TALYS cross sections ($\sigma_{\rm TALYS}$) and with cross sections ten times larger ($10\times \sigma_{\rm TALYS}$) to roughly account for these uncertainties.

\section{Results}
\label{sec: results}

\paragraph{Nucleosynthesis calculation} We use the nuclear reaction network code PRISM (Portable Routines for Integrated nucleoSynthesis Modeling) \cite{Matt2016, Matt2017, Matt} to perform the \textit{r}-process nucleosynthesis calculations to obtain the abundance patterns for the initial \textit{r}-process nuclei ejected from a NSM. Details of our baseline nucleosynthesis calculation set, are found in \cite{Wang2019}. To investigate the effects of nuclear physics variations on our spallation results, we also adopt $\beta$ decay rates of \cite{MT} and neutron capture rates from NONSMOKER \cite{NS1,NS2,NS3}\footnote{https://nucastro.org/nonsmoker.html}, in addition to the baseline neutron capture rates calculated by the Los Alamos National Laboratory (LANL) statistical Hauser-Feshbach code of \cite{HF} and baseline $\beta$ decay rates from \cite{Moller2003}.
We adopt two kinds of NSM trajectories to compare astrophysical conditions: cold dynamical ejecta \cite{Goriely2011, Matt} and a low entropy accretion disk wind which is parameterized similar to conditions in \cite{McLaughlin2005, Surman2006, Just2015, Martin2015, Wanajo2014, Siegel2018}. 

Spallation effects on \textit{r}-process nuclei ejected from a NSM depend both on the astrophysics conditions and the nuclear physics inputs for the nucleosynthesis calculations and the propagation process. In this work, we examine the effects of spallation on the $A\sim 196$ peak (third peak) region of \textit{r}-process abundance patterns for disk wind and dynamical ejecta, and we explore how the predicted influence of spallation varies with the input nuclear physics. A more complete study, including spallation of $A\sim 130$ peak nuclei and a full NSM simulation, is described in \cite{Wang2019}.

\subsection{Variations in Astrophysical Conditions}
\label{subsec: astro}

\begin{figure}[]
%\centering
\includegraphics[scale=1]{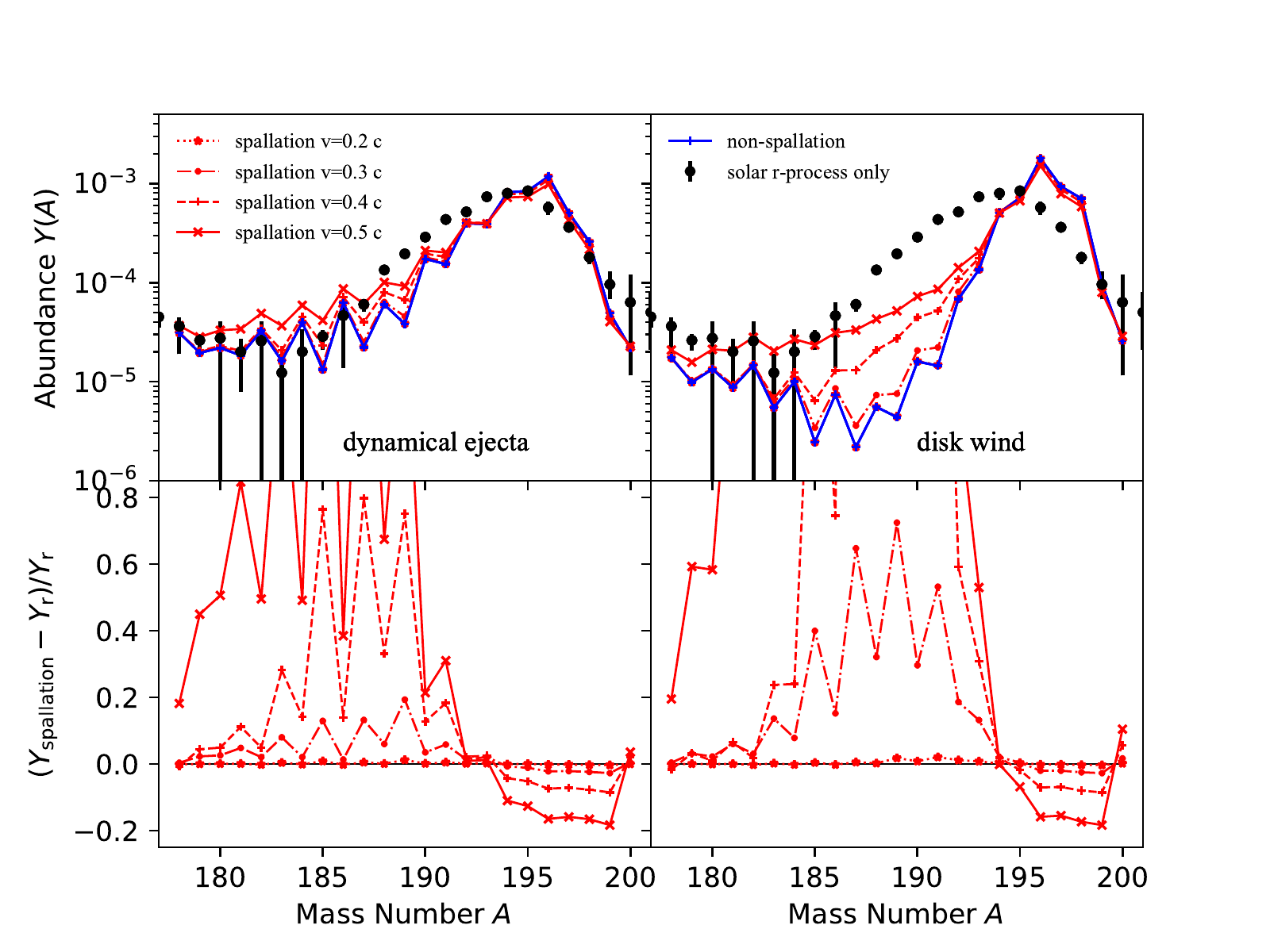}
\caption{
Spallation effects in the third peak ($180<A<200$) region for the baseline disk wind (left) and dynamical ejecta (right) simulations assuming initial ejecta velocities of $0.2c$, $0.3c$, $0.4c$ and $0.5c$. The initial \textit{r}-process abundance pattern from the PRISM simulation is shown in blue and the black points are the solar \textit{r}-process residuals \cite{solar2007}. The solar data scales to the $^{195}$Pt abundance from the initial \textit{r}-process simulation. Upper panels: Abundances before (blue lines) and after (red/orange lines) spallation. Lower panels: The abundance change ratio due to spallation as defined in \Eref{eq:change_ratio}.
%The initial \textit{r}-process abundance pattern from the PRISM simulation is shown in blue and the black points are the solar \textit{r}-process residuals \cite{solar2007}. The solar data scales to the $^{195}$Pt abundance from the initial \textit{r}-process simulation. Upper panels: Abundances before (blue lines) and after (red/orange lines) spallation. Lower panels: The abundance change ratio due to spallation as defined in Eq~\ref{eq:change_ratio}.
\label{fig:spal_v3}
}
\end{figure}

We first explore spallation effects on our baseline \textit{r}-process abundance pattern with different astrophysical conditions: different astrophysical trajectories and different initial velocities. 

%Figure~\ref{fig:spal_v} shows the resulting abundances of the third \textit{r}-process peaks before and after spallation calculations for both the hot disk wind (left) and dynamical ejecta (right) with baseline calculations.

Figure~\ref{fig:spal_v3} compares the abundance patterns and abundance change ratios in the third \textit{r}-process peaks after spallation for the baseline cold dynamical ejecta (right) and hot disk wind conditions (left). We can see that, for both trajectories, spallation moves the \textit{r}-process abundance pattern to lower mass numbers, towards the solar data, and smooths the shapes at the left side of the peaks while leaving the right side of the peaks largely unchanged. 
In addition, hotter \textit{r}-process freeze-out conditions are characterized by more late-time neutron capture, which produces abundance peaks that can be narrower than and offset from solar data, as shown in the blue lines of Figure~\ref{fig:spal_v3}. Thus, the effects of spallation are much larger with the sharper peak; the average positive spallation abundance change is $\sim200\%$ for the wind example versus $\sim40\%$ for the cold dynamical ejecta example with an initial velocity of $0.4c$. {\em Spallation effects are bigger for steeper abundance features}.

Figure~\ref{fig:spal_v3} also shows the spallation effects with different initial ejecta speeds, varying from $0.2c$ to $0.5c$ for both the dynamical ejecta and disk wind conditions. We can see that {\em the influence of spallation strongly depends on the initial velocity of the \textit{r}-process ejecta}, and the abundance pattern changes are non-negligible for ejecta of $0.3c$ or faster. For the dynamical ejecta as an example, at $0.3c$, spallation brings an $\sim 8\%$ on average of the abundance change; at $0.5c$, the abundance change can be as high as a factor of 2 for some nuclei.

\subsection{Variations in Input Nuclear Physics}
\label{subsec: nuclear}

\begin{figure}[]
%\centering
\includegraphics[scale=1]{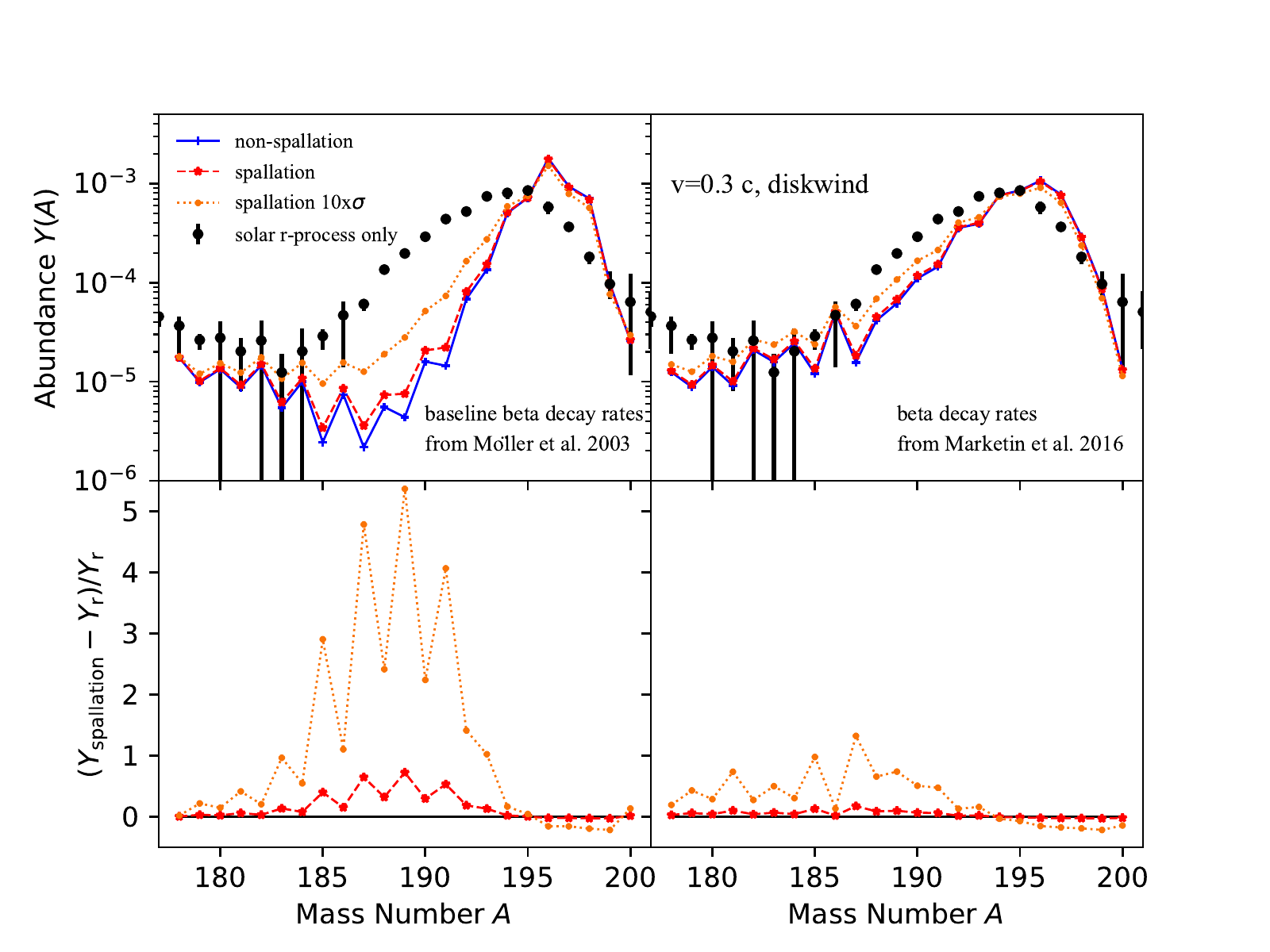}
\caption{
Spallation effects in the third peak ($180<A<200$) region for the baseline hot disk wind simulations assuming initial ejecta velocity of $0.3c$, with two choices of $\beta$ decay rates (Left: baseline calculation rates from \cite{Moller2003}; Right: rates from \cite{MT}) and two choices of spallation cross sections, $\sigma_{\rm TALYS}$ (red dashed line) and $10\times \sigma_{\rm TALYS}$ (orange dotted line). Upper panels: Abundances before (blue lines) and after (red/orange lines) spallation, compared to solar data as in Figure~\ref{fig:spal_v3}. Lower panels: The abundance change ratio due to spallation as defined in \Eref{eq:change_ratio}.
\label{fig:spal_r3}
}
\end{figure}

In Section~\ref{subsec: astro}, we have considered \textit{r}-process ejecta traveling through the ISM with different (but still uniform) initial velocities and different astrophysical conditions for the r-process nucleosynthesis calculations. Here we repeat the analysis of Section~\ref{subsec: astro} with different choices of nuclear physics adopted for the nucleosynthesis and propagation process, while keeping the initial velocity of the ejecta at $v=0.3c$ for the same disk wind trajectory.

As discussed in Sections~\ref{subsec: cross}, different theoretical calculations result in different spallation cross-section values, thus affecting the spallation abundance changes based on \Eref{eq: fraction}. Furthermore,  the \textit{r}-process proceeds through a region of the nuclear chart where the nuclear properties are highly uncertain \cite{Matt2016}. Different choices of nuclear data yield different initial \textit{r}-process patterns. Therefore the choice of nuclear data in our calculation leads to a variance in the potential influence of spallation.

Figure~\ref{fig:spal_r3} compares the abundance patterns and abundance change ratios after spallation in the third \textit{r}-process peaks for the baseline hot disk wind conditions with $\beta$ decay rates from \cite{Moller2003} (baseline nuclear rates; left) and from \cite{MT} (right), which assume an initial ejecta velocity of $0.3c$, and spallation cross sections from TALYS ($\sigma_{\rm TALYS}$; red line) and $10\times \sigma_{\rm TALYS}$ (orange line). 
We can see that {\em the spallation effect increases significantly with an increased spallation cross section}. Compared with baseline rates, $\beta$ decay rates from \cite{MT} (MKT) act to broaden the third peak and move the peak position towards solar, resulting in a flatter abundance shape and smaller abundance changes due to spallation (Baseline: $\sim30\%$ in average; MKT:  $\sim8\%$ in average).  
We also repeat our spallation calculation starting with abundance patterns produced with neutron capture rates calculated with NONSMOKER \cite{NS1,NS2,NS3}. which also bring a broader and flatter third peak and thus a smaller spallation effect ($\sim20\%$ in average) than baseline calculation.

\section{Discussions and Conclusions}
\label{sec: discussion}

In this paper, we present a thick-target spallation model to investigate how the \textit{r}-process abundance pattern produced in fast NSM ejecta is influenced by spallation reactions with ISM nuclei. We find spallation to have non-negligible effects on relative abundances in the $A\sim 130$ and $A\sim 196$ \textit{r}-process peak regions for material ejected with speeds above $0.2c$. The effects of spallation are to move the peak abundances towards lower mass numbers and to smooth the slope of the left side of the peaks. The extent to which spallation reactions can reshape the \textit{r}-process peaks depends on the relevant spallation cross sections, the initial abundance pattern and the initial bulk velocity of the ejecta, and here we explore different astrophysical conditions of the r-process ejecta and different choices of input nuclear physics. We find in cases where the initial abundance peaks are sharper or offset from solar data, spallation can partially or fully alleviate the mismatch.

Our work calls for new measurements of spallation reactions of \textit{r}-process heavy nuclei at energy range $\sim5-100$ MeV, which will test the importance of spallation in shaping NSM \textit{r}-process abundance patterns. The most important spallation targets are $ ^{198}$Pt, $ ^{197}$Au, $ ^{196}$Pt and $ ^{195}$Pt, based on our sensitivity study of the spallation cross-sections \cite{Wang2019}. These measurements could be within reach for appropriate facilities such as FRIB, FAIR and RIKEN.

\ack
We are very grateful to Luke Bovard for providing us trajectories from his neutron star merger simulations. 
We are pleased to thank George Fuller, Wick Haxton, and Shunsaku Horiuchi for the stimulating conversations.  
This work was supported by U.S. National Science Foundation under grant number PHY-1630782 Focused Research Hub in Theoretical Physics: Network for Neutrinos, Nuclear Astrophysics, and Symmetries (N3AS) (X.W.), and was also supported by the U.S. Department of Energy under Nuclear Theory Contract No. DE-FG02-95-ER40934 (R.S.), DE-AC52-07NA27344 for the topical collaboration Fission In \textit{R}-process Elements (FIRE; N.V. and R.S.),  and the SciDAC collaborations TEAMS DE-SC0018232 (T.S., R.S.).  
This work benefited from conversations stimulated by the National Science Foundation under Grant No. PHY-1430152 (JINA Center for the Evolution of the Elements). 
M.M. was supported by the US Department of Energy through the Los Alamos National Laboratory. Los Alamos National Laboratory is operated by Triad National Security, LLC, for the National Nuclear Security Administration of U.S.\ Department of Energy (Contract No.\ 89233218CNA000001). T.S. was supported in part by the Los Alamos National Laboratory Center for Space and Earth Science, which is funded by its Laboratory Directed Research and Development program under project number 20180475DR. M.M. was also supported by the Laboratory Directed Research and Development program of Los Alamos National Laboratory under project number 20190021DR.

%\appendix

\end{document}